\documentclass[12pt,preprint]{aastex}
\usepackage{subfigure}

\slugcomment{To appear in ApJ, hopefully}

\shorttitle{Chandra observations of red quasars}
\shortauthors{Urrutia et al.}

\begin{document}

\title{Chandra Observations of 12 Luminous Red Quasars}

\author{Tanya Urrutia\altaffilmark{1,2}, 
Mark Lacy \altaffilmark{1,2,3}, 
Michael D.\ Gregg \altaffilmark{1,2},
Robert H.\ Becker \altaffilmark{1,2}}

\altaffiltext{1}{Department of Physics, University 
of California, One Shields Avenue, Davis, CA 95616; 
urrutia@physics.ucdavis.edu }

\altaffiltext{2}{IGPP, L-413, Lawrence Livermore National Laboratory, 
Livermore, CA 94550; bob@igpp.ucllnl.org, gregg@igpp.ucllnl.org}

\altaffiltext{3}{Spitzer Science Center, MS 314-6, California Institute of 
Technology, 1200 E.\ California Boulevard, Pasadena, CA 91125; 
mlacy@ipac.caltech.edu}

\begin{abstract}
We present results of a study of 12 dust-reddened quasars with 0.4 $<$ z $<$ 
2.65 and reddenings in the range 0.15 $<$ $E(B-V)$ $<$ 1.7. We obtained 
ACIS-S X-ray spectra of these quasars, estimated the column densities towards 
them, and hence obtained the gas:dust ratios in the material obscuring the 
quasar. We detect all but one of the red quasars in the X-rays. Even though 
there is no obvious correlation between the X-ray determined column densities 
of our sources and their optical color or reddening, all of the sources show 
absorbed X-ray spectra. When we correct the luminosity for absorption, they 
can be placed among luminous quasars; therefore our objects belong to the 
group of high luminosity analogues of the sources contributing to the X-ray 
background seen in deep X-ray observations. Such sources are also found in 
serendipitous shallow X-ray surveys. There is a hint that the mean spectral 
slope of the red quasar is higher than that of normal, unobscured quasars, 
which could be an indication for higher accretion rates and/or an 
evolutionary effect. We investigate the number density of these sources 
compared to type 2 AGN based on the X-ray background and estimate how many 
moderate luminosity red quasars may be found in deep X-ray fields.
\end{abstract}


\section{Introduction}

For a long time, deciphering the nature of the X-ray background (XRB) was a 
key problem in astrophysics. In the last years with the launch of the new 
generation of X-ray telescopes, {\it Chandra} and {\it XMM/Newton}, it 
has been possible to resolve the XRB into discrete point sources in very deep 
X-ray images (e.g. Chandra Deep Field South \citep{cdfs1}, Lockmann Hole 
\citep{lockmann}, Chandra Deep Field North \citep{cdfn1}). Most of these 
point sources have harder spectra than known quasars. About 90\% of the X-ray 
sources in these deep fields have an optical counterpart with R $\lesssim$ 
24 \citep{cdfn2}. Optical identification of these sources shows that while 
some objects are normal quasars, many are reddened quasars, narrow-line AGN 
or low redshift ``normal'' galaxies \citep{cdfs2}. As a complement to the 
deep X-ray fields, studies of X-ray weak quasars have yielded harder X-ray 
spectra and redder optical colors than normal quasars \citep{risa03}.

The discovery of large numbers of reddened quasars and narrow-line AGN 
implies that a large amount of the accretion luminosity from material falling 
into black holes is hidden by dust (in the optical) and gas (in the soft 
X-ray). Although there is order-of magnitude agreement between the amount of 
radiation produced by quasars over the lifetime of the Universe and the mass 
accreted onto the black holes we see in the centers of galaxies today 
\citep{mf01}, there is still discussion on the fraction of quasars that are 
obscured by dust. Population synthesis analysis of the hard XRB estimate that 
up to 90\% of AGN, whose evolution peaks at z $\approx$ 0.7, are absorbed by 
dust \citep{gsh,ueda}. This fraction of obscured quasars does not coincide 
with the quasar population we are used to seeing in the optical, where the 
majority of quasars are blue, unabsorbed and peak at larger redshifts. We 
might be missing a large number of dust reddened quasars \citep{webster}. It 
remains to be investigated what fraction of the quasar population we are 
missing in optical color-selected surveys, which are most effective at 
selecting quasars with blue optical colors.  

In addition to the X-ray surveys, various techniques have been developed to 
find quasars that are redder than those typically found in optical quasars 
surveys. Radio selected quasars overall seem to have redder colors 
\citep{francis,reddest,postman} than optical selected samples, although there 
is still some debate about how much of this redness can be produced by 
optical synchrotron emission in these powerful radio sources. In addition, 
broad absorption line quasars (BAL quasars) samples have redder colors than 
typical quasars \citep{sp92,reich} and higher X-ray column densities 
\citep{gal} also suggesting that a large fraction of this population is 
missing from optical selected samples. Searches for red quasars have yielded 
a higher than normal amount of lensed systems \citep{reddest,reddest2}, whose 
discovery is aided by magnification bias, raising the possibility of large 
numbers of unmagnified obscured objects.

Although it seems clear that the reddening of the quasars is caused by the 
optical absorption by the warm dust obscuring the nuclear region, we do not 
yet know where this obscuration is taking place. The dust could be located in 
the host galaxy's ISM, but obscuration could also arise due to an accretion 
disk's dusty regions. There is evidence that dust exists in the Narrow Line 
Region (NLR) of AGN \citep{radom} and even as we move inward of the accretion 
disk large dust grains could survive the radiation field in dense molecular 
clouds \citep{dopita}. The location of the dust responsible for reddening is 
important in that we can compare it with the obscuring gas in the X-rays. 
\cite{mai01} speculate that AGN classification in the optical is different 
than in the X-ray with $E_{B-V}$/$N_H$ ratios lower than Galactic by a factor 
of $\sim$ 3 up to $\sim$ 100. We do not know what this ratio is for extremely 
absorbed quasars. Simple AGN models (e.g. \cite{anton}) cannot explain this 
high dispersion, so the nuclear region of AGN is apparently more complicated 
than usually assumed. 

In this paper we present results of X-ray observations of 12 red quasars with 
{\it Chandra} and compare these results with other X-ray observation of red 
AGN \citep{wilkes}. Our principal aim is to investigate whether our optically 
selected are the high luminosity analogues of the faint red quasars found in 
the deep Chandra fields. Other shallow serendipitous X-ray surveys have found 
cases of high luminosity, obscured or red Type 2 quasars \citep{fiore}, so 
our goal is that our objects belong to those group of objects fall into this 
category. We adopt a flat universe, $H_0 = 72 \,km \,s^{-1}/Mpc$, 
$\Omega_{\Lambda} = 0.7$ cosmology.

\section{Observations and data analysis}

\subsection{Optical selection process}

To find red quasars which could have been missed in optical surveys we
applied the following three criteria: (1) the object has to be as FIRST
radio source \citep{first}, (2) it must be in the 2MASS point-source catalog 
\citep{2mass}, and (3) it must have $R-K\stackrel{>}{_{\sim}}4.8$ based on 
comparison with the Palomar Observatory Sky Survey first (POSS-I) or second 
generation (POSS-II) images \citep{apm}. These criteria are similar to those 
of \cite{f2m}, but differ in one significant respect in that we also include 
objects which are detected on the POSS-I as long as they satisfy our 
$R-K\stackrel{>}{_{\sim}}4.8$ criterion (using the POSS E magnitude as an 
approximation for R), while Glikman et al. did not. This results in the 
inclusion of several objects which are not present in the sample of 
\cite{f2m}. 

The resulting candidate list included 20 objects detected on the POSS-I 
plates in addition to the 69 undetected objects in the Glikman et al. 
candidate list. Optical magnitudes for the candidates were obtained from the 
POSS-II plates, or from the Sloan Digital Sky Survey (SDSS) if they were in 
the area covered by data release 2 \citep{sdss-dr2}. Irregular host galaxies 
of or companions to our lower-redshift red quasars can often be seen in the 
SDSS data, in agreement with studies of other red quasars \citep{hines}. 

Spectroscopy of our candidate quasars was carried out in the optical and 
near-infrared at various facilities. Objects with emission line widths 
$>$1000km $s^{-1}$ were classified as quasars. 28 of our 89 candidates 
turned out to be broad-line quasars with extremely red spectral energy 
distributions. The spectra of seven of our objects can be seen in \cite{f2m}, 
\cite{reddest} and \cite{reddest2}. The remaining five objects have similar 
spectra (red continuum, broad quasar lines, no features associated with 
stellar light). Our requirement for detection of broad lines naturally 
excludes Type 2 AGN, and thus results in an expected X-ray column density 
$N_{H} < 10^{24} {\rm cm^{-2}}$. 

The selection of red quasars based on $R-K$ colors can, in principle, include 
objects which are red due to galaxy starlight, hence the need for high 
quality spectra. However, since all the objects under discussion in this 
paper are high luminosity quasars, it is not possible that the colors are 
dominated by starlight. Nonetheless, a quantitative discussion must go beyond 
colors to an actual determination of $E(B-V)$. Where our spectra covers both 
$H{\alpha}$ and $H{\beta}$, the Balmer decrement can be used for this 
purpose. This is the case for nine of the twelve objects. In calculating the 
Balmer decrements, we assumed an intrinsic $H\alpha / H\beta$ = 3.2 from the 
SDSS quasar composite \citep{sdssq}. The errors from the Balmer decrement 
measurements are derived from propagating the line flux errors. Since the 
broad Balmer lines are presumed to arise only in quasars, the reddening 
derived from the Balmer decrement is independent of any starlight in the 
spectrum. In the three spectra, which didn't cover both $H{\alpha}$ and 
$H{\beta}$ (FTM0906$+$4952, FTM0915$+$2418 and FTM1036$+$2828) we have fit 
the continuum slope following the technique described in \cite{f2m}. The 
spectra of these three quasars show no sign of starlight, e.g. from stellar 
absorption features (nor do the other nine spectra). Furthermore, the 
continuum estimates ranged up to the K-band, where in the case of our 
luminous obscured quasars, starlight has a negligible contribution to the 
spectrum, so the continuum slope should give a meaningful estimate of $E(B-V)$.

It turns out that FTM0906$+$4952 was a variable source which was especially 
bright when the infrared observations (2MASS) were being taken. Its reddening 
($E(B-V)$ = 0.15) is still is somewhat redder than the mean SDSS quasar, but 
it does not compare to the extreme extinction shown by our other quasars. 
Nevertheless, we opted to keep FTM0906$+$4952 in our sample for comparison 
reasons. $E(B-V)$ values are quoted in Table \ref{observe}.

Our sample for study with Chandra was restricted to z $>$ 0.4 to ensure that 
it contained moderate to high luminosity quasars (the least well-studied 
obscured AGN population in the X-ray). Also, to ensure that the red colors of 
our quasars were not caused by an optical synchrotron component 
\citep{wwf01}, we included only objects with faint radio fluxes ($<20$mJy at 
1.4GHz), or objects with higher radio fluxes but which show clear Balmer 
decrements in their optical/near-infrared spectra. Among the quasars chosen 
are FTM0134$-$0931 (a gravitational lens) and FTM0738$+$2750, two sources 
already discussed in \cite{reddest}.

\subsection{X-ray observations}

We followed up on 12 of the red quasars mentioned in section 2.1 in the 
X-rays to find out if these red quasars represent the high luminosity 
counterparts of the obscured quasars found in the deep Chandra fields. 
Observations were carried out with {\it Chandra}, six quasars were observed 
during AO3 and six during AO5 as part of the GO program (proposal numbers 
03700742 and 05700838). For the observations the ACIS-S backside illuminated 
CCD (S3) with a 1/8 sub-array was used (faint mode). Table \ref{observe} 
shows the journal of observation of our quasars.

The ACIS images were analyzed with {\it CIAO} package, version 3.0.1. We 
used the detection algorithm {\it celldetect}; all but one source were 
detected at the positions predicted by the optical data. The non-detected 
source (FTM0738$+$2750) displays two very hard photons at the optical 
position of the quasar. FTM1022$+$1929 shows two peaks, it might be a missed 
gravitational lens, even though the detection algorithm classified it as one 
source. 

Table \ref{ciao} shows the X-ray properties of our sources. In our convention 
0.2 to 10.0 represents the broad, 0.2 to 2.0 keV the soft and 2.0 to 10.0 keV 
the hard energy range. The hardness ratio (HR) is defined as 
$HR = (H - S)/(H + S)$, where S are the number of photons in the soft band 
and H are the number of photons in the hard band. Table \ref{hr} shows our 
deduced HRs in the first column. As these are not the bands typically used in 
Chandra surveys, we also show the HR with 0.5 - 2.0 keV in the soft band and 
2.0 - 7.0 keV in the hard band in column 2 of Table \ref{hr}. As expected, 
the color or hardness ratio of FTM0906$+$4952 (the variable source) in the 
X-ray was the softest in our sample. 

\begin{figure}
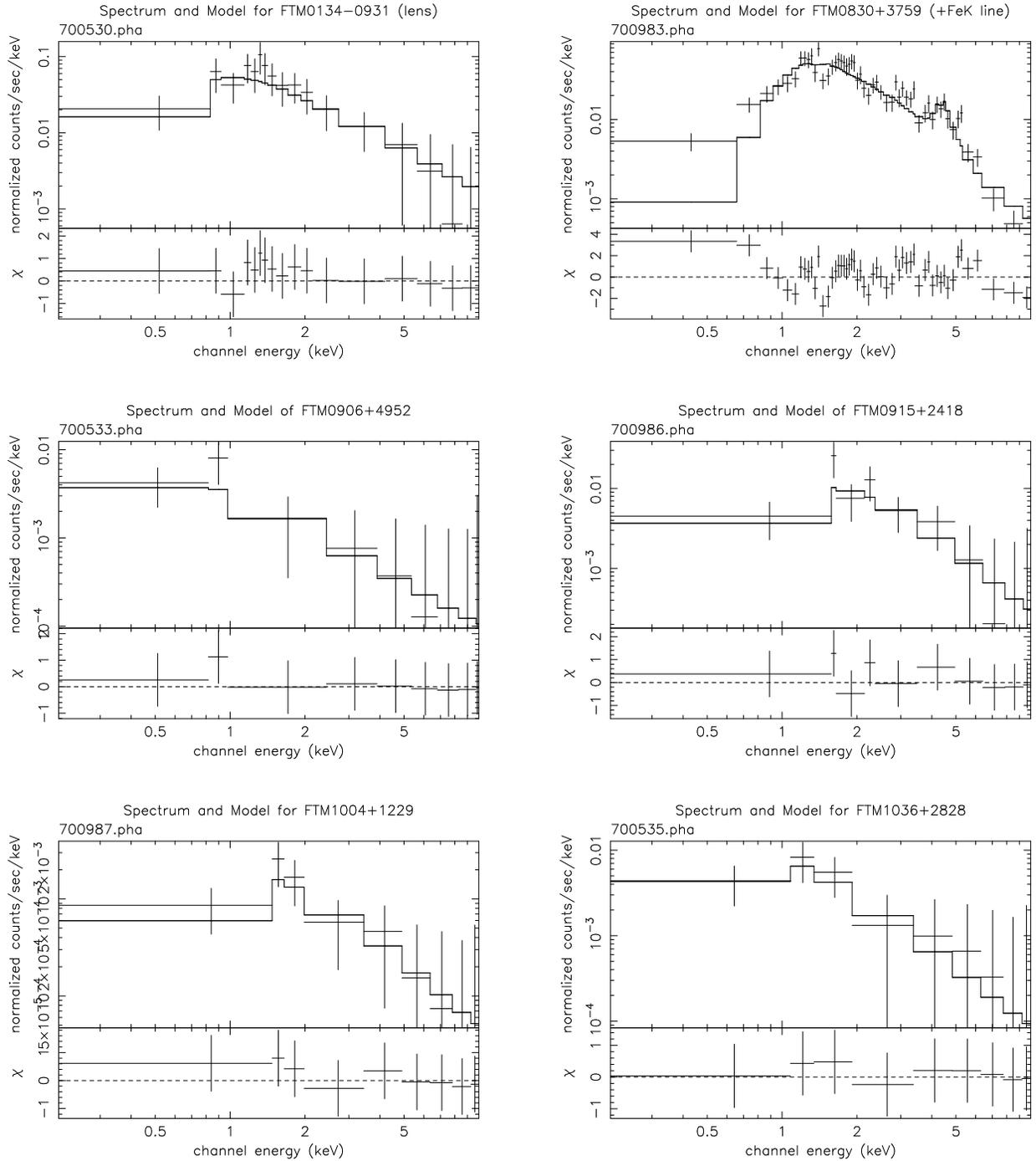

\begin{center}
\subfigure{\includegraphics[height = 7.5cm, angle=-90]
{f1a.ps}}
\hspace*{1cm}
\subfigure{\includegraphics[height = 7.5cm, angle=-90]
{f1b.ps}}
\subfigure{\includegraphics[height = 7.5cm, angle=-90]
{f1c.ps}}
\hspace*{1cm}
\subfigure{\includegraphics[height = 7.5cm, angle=-90]
{f1d.ps}}
\subfigure{\includegraphics[height = 7.5cm, angle=-90]
{f1e_new.ps}}
\hspace*{1cm}
\subfigure{\includegraphics[height = 7.5cm, angle=-90]
{f1f.ps}}
\end{center}
\caption{X-ray spectra of the sources with more than 40 counts. Shown are the 
counts with at least 2$\sigma$ significance. They were fitted with an 
absorbed power-law model. For FTM0134-0931 we show the model with the lens as 
the absorber. For FTM0830+3759 we show 4$\sigma$ significance, since the 
spectrum has over 700 counts. This spectrum also has a broadened Fe K line 
added to it. It also shows other absorption and emission features, which 
cannot be resolved. 
\label{spectra}}
\end{figure}

We then extracted the source and background pulse height amplitude (PHA) 
spectra for the X-ray sources with more than 40 counts detected. For the 
source a 5'' circle region and for the background a (30'',15'')-annulus 
region were used. The response function (RMF) was then extracted 
corresponding to the region of the chip, where the source was located. The 
spectral analysis of the X-ray sources was carried out with {\it XSPEC} 
version 11.2.0, as part of the {\it Xanadu} package, obtainable at HEASARC 
({\it http://xspec.gsfc.nasa.gov}). An absorbed power-law model at 
the quasar redshift was fitted to the spectra. We used the photo-electric, 
redshift corrected absorption model, with Wisconsin cross sections 
\citep{zwabs}, in Xspec: {\it zwabs}. These fits the yielded us the values of 
the spectral slope ($\Gamma$) and column density ($N_H$) in the rest frame of 
the quasar (see Table \ref{xspec}). The resulting spectra are shown in Figure 
\ref{spectra}.

\section{Results}

One of the remarkable, if not surprising, results we find, is that all of the 
red quasars are absorbed to some degree in the X-rays, as they have harder 
X-ray colors than typical quasars. Most of our sources have a low S/N, 
therefore we cannot get accurate spectral slopes and column densities. 
However, from the measured HRs alone, we can extract useful information for 
the faint sources. Our sample has a mean HR of 0.08 $\pm$ 0.11 (observed 
frame). The mean HR changes very little (to 0.09 +/- 0.10) when we calculate 
it with the bands normally used in Chandra deep fields: soft = 0.5 - 2.0 keV, 
hard = 2.0 - 7.0 keV (see Table \ref{hr}). This value is higher than expected 
in non-biased X-ray surveys, where the median HR of typical quasar lies more 
around -0.5 \citep{cdfs2}. We then corrected the HR for redshift. For 
absorbed sources the absorption cutoff moves to lower energies when the 
object is at high redshift, so the corrected HR tends to be softer at higher 
redshift. We chose a HR at z = 1, that is the 0.4 - 4.0 keV and 4.0 - 20.0 
keV bands in the rest frame, the values are quoted in column 3 of Table 
\ref{hr}. The mean for the corrected HR is 0.15 $\pm$ 0.10.

\begin{figure}
\begin{center}
\plotone{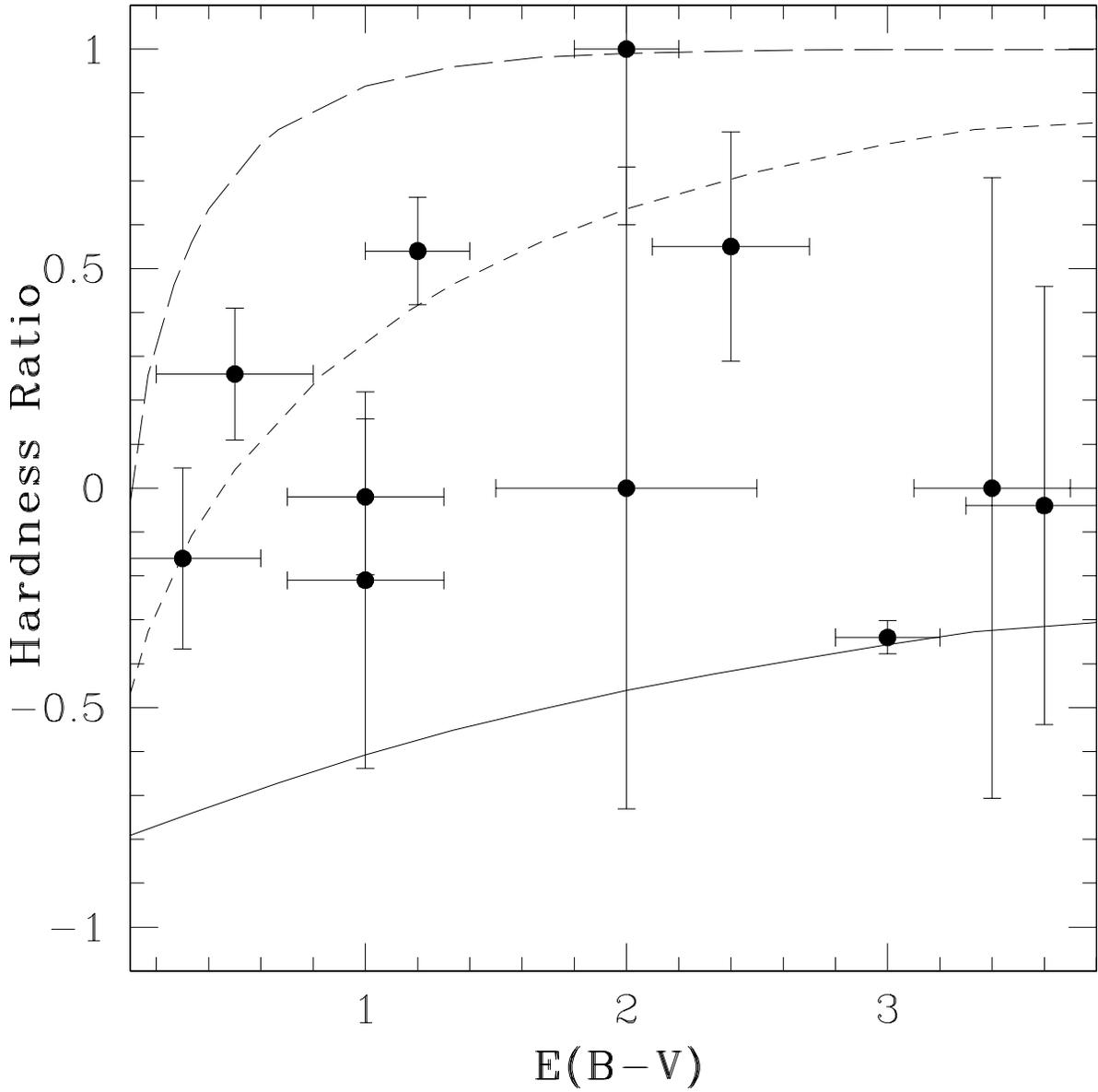}
\end{center}
\caption{Distribution of hardness ratios versus reddening. The hardness 
ratios were corrected for redshift, so that they represent a hardness ratio 
at z=1. The solid line is a model with Galactic gas:dust ratio, for the small 
dotted line the gas:dust ratio is 20 times and for the long dashed line 100 
times Galactic value.
\label{hr-col}}
\end{figure}

Figure \ref{hr-col} shows the distribution of redshift corrected hardness 
ratios versus reddening. There is no clear correlation between the reddening 
and the hardness ratio of our sample. Shown are 1$\sigma$ error bars on the 
HR to get an idea how low the S/N is. We have calculated quasar models with 
$\Gamma = 2$ and different gas:dust ratios. The solid line represents the a 
Galactic gas:dust model ($E (B-V) = N_H / 6 \times 10^{21} cm^{-2}$), the 
small dashed line represents a model with 20 times Galactic gas:dust ratio 
and the large dashed line is a model with a gas:dust ratio of 100 times 
Galactic value. 

Wilkes' X-ray observations of a sample of 26 low redshift AGN \citep{wilkes},
has a mean observed frame HR of +0.14 $\pm$ 0.05, slightly higher than our 
our sample. This higher HR is expected for lower redshift models as the 
absorption cutoff is at higher energies. Again, for the few objects of 
Wilkes' sample for which we could obtain X-ray spectral information, no 
obvious correlation between reddening and column density is found.

Nevertheless, our redshift corrected HRs all lie around the regime where the 
column density tends to be around a few times $10^{22} cm^{-2}$, so there 
should be obscuration in the quasars, although not the obscuration we 
typically associate with Seyfert 2 galaxies, where the typical column density 
lies around $10^{24} cm^{-2}$. Column densities around $10^{22} cm^{-2}$ are 
directly verified in those cases having large enough counts to allow a proper 
spectral fit (Table \ref{xspec}), which seems to indicate that the absorbed 
power law model we chose for the spectra is the correct one. 

Also, the fact that our points all have column densities larger than expected 
from the dust reddening (assuming a Galactic gas:dust ratio) confirms the 
results of \cite{mai01}, who claim that the $E(B - V) / N_H$ ratio is lower 
than Galactic by a factor of $\sim$3 to 100 and that this is due to the 
circumnuclear region being dominated by large dust grains. For the largest 
reddenings in Figure \ref{hr-col}, though, the gas:dust ratio seems to 
approach Galactic value, in contradiction to \cite{mai01}. One possible 
explanation for this is that the ISM of the host galaxy itself could be doing 
the obscuration for the extremely reddened objects.

For the quasars, where we could extract a spectrum we corrected for the 
absorption given by our models deduced with XSPEC (Table \ref{xspec}). The 
corrected luminosity value for FTM0134$-$0931 is likely to be wrong as there 
is luminosity enhancement due to the gravitational lens, but the correction 
for the other quasars places them among the high section of the quasars 
luminosity function. Our quasars therefore should represent only the tip of 
the red quasar iceberg, at lower X-ray luminosities we should find higher 
numbers of them, which is in agreement with the results from the X-ray deep 
fields. Luminosity correction for the AGN in Wilkes' sample do not show such 
high luminosities, a logical finding, as they are at lower redshifts. The 
fact that we don't see these exceedingly large luminosities at low redshift 
in the Wilkes sample rules out the possibility that all highly reddened 
quasars are highly luminous. As our sample is highly luminous and at higher 
redshift, we are obviously missing an even larger, as yet undiscovered, 
population of quasars at lower luminosity, in agreement with findings from 
\cite{postman}.

It has often been put forward that the higher than normal steepness of the 
X-ray spectrum in BAL quasars and narrow line Seyfert 1 galaxies (NLS1) is a 
signature of a high accretion rate \citep{mathur2000,boller}. \cite{bobbal} 
speculate that the popular notion that all BAL quasars are normal quasars 
seen edge-on is wrong, but that the nature of BALs could be more an 
evolutionary effect. If BAL quasars and NLS1s were to represent an early 
phase in quasar evolution, the steeper than normal spectrum we find in red 
quasars, could also represent an evolutionary phase in the lifetime of an 
quasar. To investigate this possibility, we need a bright enough sample to 
find the intrinsic power law, as the quasars themselves are obscured. We have 
to probe this hypothesis on the quasars that lie on the high end of the 
intrinsic X-ray luminosity function. 

\begin{figure}
\begin{center}
\plotone{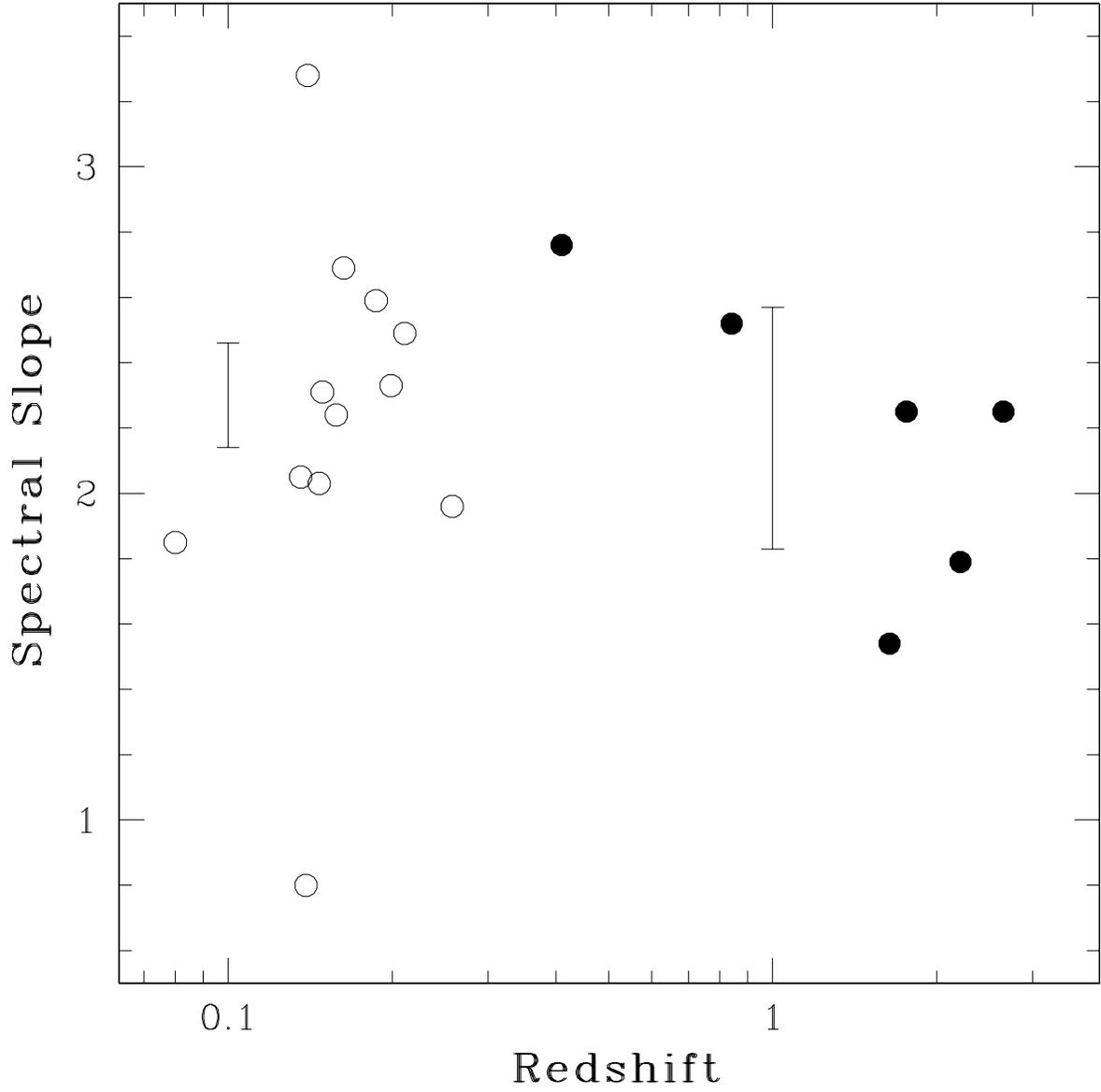}
\end{center}
\caption{Spectral slopes vs. redshift. Filled dots are from our sample, open 
dots are from Wilkes' low redshift red AGN sample. The bars show the typical 
errors in each of the samples.
\label{gamz}}
\end{figure}

The quasars with spectral information also present a somewhat steeper than 
normal spectral slope than normal quasars. The unweighted mean $\Gamma$ for 
our sample is 2.2 $\pm$ 0.4. We use the unweighted mean, because otherwise 
we would be dominated by FTM0830$+$3759, which has very small errors 
(weighted mean $\mu$' = 2.8 $\pm$ 0.1). The fact that our red quasars are 
radio selected actually contradicts broad band studies of quasars that have 
found that radio-loud quasars tend to have flatter spectra in the X-rays 
($\Gamma \approx 1.6$) than radio-quiet quasars ($\Gamma \approx 1.8$), 
although there is a large scatter for both types \citep{elvis}. Usually 
Seyfert 1 galaxies tend to have $<\Gamma>$ 1.8 - 1.9 for energies between 
0.2 - 10.0 keV \citep{walterfink}, so our mean is slightly above the expected 
spectral index. To make sure that we are getting the correct spectral slopes 
we also fit a simple power-law to the high energy region in the spectra, 
where absorption is not a factor. These new spectral slopes do not vary from 
the other ones significantly and are well within the errors. Their mean is 
2.1 $\pm$ 0.5. The errors are larger, since we are fitting even fewer counts. 
The fact that the slopes from the fit to the high energy portion of the 
spectra are similar to those obtained from modeling of the whole spectrum 
suggests that our fitting is reliable, and that the higher spectral slopes we 
obtain are not an artifact of the absorbed power law model we chose. There is 
only a marginal evolution of the spectral index with redshift (Figure 
\ref{gamz}). The spectral index flattens somewhat at higher redshift, because 
the reflection hump in the rest-frame range of 10-40 keV moves into the 
Chandra band at high redshift \citep{schartel}. 

Wilkes' red AGN also have statistically larger than usual spectral index, 
with a weighted mean of 2.3 $\pm$ 0.2 and an unweighted mean of 2.2 $\pm$ 
0.2. The fact that the spectral indices for red quasars are larger than 
expected, could be an indicator of these objects having larger accretion 
rates. Just like NLS1 galaxies or BAL quasars, the red quasar phenomenon 
could be an evolutionary stage in the quasar lifetime. The quasar would be 
obscured as it has just ignited and the dust is still settling in. All of the 
deduced values of $\Gamma$ for the red quasars are still normal range for 
Seyfert 1 galaxies within the errors, but overall the values are higher. 

We now comment briefly on our two brightest sources, for which we could get 
most spectral information.

\subsection{FTM0134-0931}

This FIRST-2MASS object is extremely red (R - K = 7.61). It has been 
identified as a lensed quasar, showing absorption lines from the lensing 
galaxy in the optical spectrum \citep{reddest}. The quasar lies at a redshift 
of z = 2.21, while the lensing galaxy is at z = 0.76. Even with such a short 
exposure time of 1.1 ksec, the quasar had around 95 detection counts, which 
was enough to get spectral information. When we analyze the X-ray spectrum, 
neither an absorbed power law fit at the redshift of the quasar nor at the 
redshift of the lens seems to yield a satisfying fit. Nevertheless the 
inferred column density is strongly dependent on which system is doing the 
absorption ($N_H = 1.378 \times 10^{22} cm^{-2}$ if the lens is absorbing, 
$N_H = 5.925 \times 10^{22} cm^{-2}$ for absorption within the quasar).

One fact that complicates this analysis is the fact that in the radio, we see 
a lensing geometry of 5 radio components, it seems the lensing galaxy itself 
has a radio counterpart \citep{reddest}. It might well be that the lensing 
galaxy is an AGN, accounting for a soft component in the X-rays. So far, no 
simple model has been able to account for this object in any wavelength 
range. Further X-ray spectroscopy with a longer exposure time could support 
the claim that the lensing galaxy is an AGN if we do see hints for two 
power-laws.

\subsection{FTM0830+3759}

When FTM0830$+$3759 was observed for aprox. 9.2 ksec, it showed a bright 
X-ray source with close to 800 counts. We have fitted an absorbed power law 
to FTM0830$+$3759, which is absorbed in the soft X-rays (see Table 
\ref{ciao}). Notice that, even though this is fairly bright X-ray quasar, we 
are able to fit the broadened Iron K$\alpha$ line, which you can see from the 
``hump'' at around 4.5 keV in middle left of Figure \ref{spectra}. More 
features are clearly visible in the spectrum, although it is not clear if 
they are emission lines which become visible as the power law gets absorbed 
or if they are absorption lines associated with the warm absorbing gas.

In FTM0830$+$3759, we are probably looking almost directly into the accretion 
disk, as we see the Fe~K$\alpha$ emission line (6.4~keV) with a 4$\sigma$ 
significance. Fits with X-spec estimate its width to about 0.6 keV (4 
$\times 10^4$ km/s), which should only possible at the extreme relativistic 
areas of the accretion disk (about 20 Schwarzschild radii). This fit should, 
however be taken with a grain of salt, as it has big errors (0.3 keV). 
Better resolution will tell us what the conditions are near the black hole. It 
is interesting to note, that if we correct the luminosity for the obscuration 
obtained with the spectral fit ($2.661 \times 10^{22} cm^{-2}$), it places 
FTM0830$+$3759 among luminous X-ray quasars in the sky 
($L_x = 3.89 \times 10^{45}$ erg/s). At such luminosities, the accretion disk 
could be completely ionized and would not produce the Fe K reflection 
component, which comes from a cooler disk (``X-ray Baldwin effect'', 
\cite{nandra}). 

FTM0830$+$3759 also has the steepest spectrum of all our red quasars, a hint 
for a large accretion rate. Note from the spectrum, that the quasar probably 
has a soft excess which is not included in the model. If we ignore the 
spectrum to an energy of approx. 0.85 keV the statistics significantly get 
better (almost all $\chi < 2$), we obtain more absorption and a steeper 
spectrum ($N_H = 3.09 \times 10^{22}$ and $\Gamma = 3.1$). This further 
supports our view that FTM0830$+$3759 is a young quasar with a high accretion 
rate, an obscured analog to NLS1s \citep{mathur}.

\section{Discussion}

Even though all our quasars show absorption in the X-rays it is not enough to 
classify them as Type 2 AGN in the X-rays ($N_H > 10^{24} cm^{-2}$). This 
fact is interesting, because it is believed that highly obscured objects, 
preferably Seyfert 2 galaxies at low redshifts, make up the hard XRB, which 
peaks at around 40 keV. But so far more quasars with only moderate 
obscuration have been found in faint X-ray surveys than Type 2 quasars. 
Highly obscured quasars have been found in X-ray surveys \citep{norman}, 
however not in the numbers necessary to account for the hard XRB 
\citep{fabian}. It is still not certain what fraction of Compton-thick AGN 
make up the XRB. In that context, we don't know what fraction of the XRB is 
made up by red quasars, but our objects appear to be the high luminosity 
equivalents of the bulk faint X-ray sources, which aren't Compton thick. 

In recent times, in the Chandra deep fields, a new population of sources has 
been discovered, which have extreme X-ray/optical ratios, the so-called EXOs 
\citep{exos}. In that sense, our red quasars are the high luminosity 
equivalent of these objects, having high X-ray and IR flux compared to the 
optical. The higher accretion rates suggested by the spectral indices in our 
objects could be a hint that they are the same population. The EXO 
population, seems to have atypically higher accretion rates than implied by 
the $M_{BH}-\sigma$ when comparing the their X-ray flux to the K-band 
magnitude of their hosts. The high accretion rates are indications of our 
objects and EXOs being still young and where the dust is just settling in. 
This may be a common phenomenon, nevertheless it still remains to be seen 
what fraction of objects in deep X-ray surveys belong to the population of 
EXOs. 

Deeper optical surveys are needed to detect red quasars, beyond the highly 
luminous obscured ones. By obtaining the fraction of obscured quasars at a 
certain redshift, we can then tell if the missing population of the hard XRB 
is made up primarily of Type 1 obscured AGN or Compton thick Type 2 AGN. To 
this day, this has proven difficult because selection criteria, like 
color-color properties, to find red quasars are not yet established. For 
example, light from the host galaxies will dominate over the quasar light at 
blue wavelength. Optical surveys also are magnitude limited, so it will be a 
difficult task to find the bulk of obscured quasars \citep{postman}. Another 
fact that complicates this analysis is that we found out that the red color 
in quasars is not closely correlated to the column density of X-ray absorbing 
material, which makes it difficult to speculate on X-ray properties from 
IR/optical data alone. Using the {\it Spitzer Spitzer Space Telescope} 
\citep{spitzer} we might be able to discern the true fraction of obscured to 
unobscured AGN, due to the fact that the mid-infrared emission regions of AGN 
are less obscured and we can easily distinguish the galaxies from AGN from 
the MIR SEDs alone \citep{lacy04}.

\acknowledgments
The authors wish to thank Sally Laurent-Muehleisen for her help with the 
simulations for the Chandra proposals. They acknowledge support from 
NASA/Chandra grant number G02-31245 and grant number 005700838

This work was partly performed under the auspices of the US Department of 
Energy, National Nuclear Security Administration by the University of 
California, Lawrence Livermore National Laboratory under contract No. 
W-7405-Eng-48, and partly at the Jet Propulsion Laboratory, California 
Institute of Technology, under contract with the National Aeronautics and 
Space Administration (NASA).

%\appendix

\clearpage

\clearpage

\clearpage

\clearpage

\begin{deluxetable}{cccccccccc}
\tabletypesize\scriptsize
\tablecaption{Journal of Chandra observations \label{observe}}
\tablewidth{0pt}
\tablehead{
 & & & & \colhead{Galactic $N_H$} & & & & & \colhead{Exposure} \\
\colhead{Source} & \colhead{R.A.} & \colhead{Dec.} & \colhead{z} & 
\colhead{($ \times 10^{20} cm^{-2}$)\tablenotemark{1}} & \colhead{K-mag} & 
\colhead{R-K} & \colhead{$E(B-V)$} & \colhead{Obs. Date} & \colhead{(s)}
}
\startdata
FTM0134$-$0931 & 01 34 35.60 & $-$09 31 03.00 & 2.21 & 2.73 & 13.6 & 7.6 & 
0.6 $\pm$ 0.2 & 2002 Aug 23 & 1078\\
FTM0729$+$3336 & 07 29 10.40 & $+$33 36 34.00 & 0.95 & 5.35 & 14.5 & 5.2 & 
1.0 $\pm$ 0.2 & 2002 Feb 26 & 1899\\
FTM0738$+$2750 & 07 38 20.10 & $+$27 50 45.50 & 1.99 & 4.91 & 15.3 & 5.5 & 
1.3 $\pm$ 0.4 & 2002 Apr 14 & 3998\\
FTM0830$+$3759 & 08 30 11.10 & $+$37 59 51.90 & 0.41 & 4.23 & 14.6 & 4.8 & 
1.5 $\pm$ 0.2 & 2004 Jan 26 & 9203\\
FTM0841$+$3604 & 08 41 05.00 & $+$36 04 50.10 & 0.55 & 3.41 & 14.9 & 5.9 & 
1.7 $\pm$ 0.3 & 2004 Feb 21 & 9318\\
FTM0904$-$0145 & 09 04 50.50 & $-$01 45 24.60 & 1.00 & 2.74 & 14.9 & 4.8 & 
1.0 $\pm$ 0.5 & 2003 Nov 25 & 7515\\
\vspace*{-0.1cm}\\
FTM0906$+$4952 & 09 06 51.50 & $+$49 52 36.00 & 1.64 & 1.78 & 15.1 & 6.1 & 
0.15 $\pm$ 0.3\tablenotemark{2} & 2002 Jun 01 & 5385\\
FTM0915$+$2418 & 09 15 01.70 & $+$24 18 12.20 & 0.84 & 3.84 & 13.8 & 5.5 & 
0.5 $\pm$ 0.3\tablenotemark{2} & 2003 Nov 27 & 3205\\
FTM1004$+$1229 & 10 04 24.90 & $+$12 29 22.40 & 2.65 & 3.59 & 14.5 & 6.3 & 
1.2 $\pm$ 0.3 & 2004 Jun 07 & 17666\\ 
FTM1012$+$2825 & 10 12 30.50 & $+$28 25 27.20 & 0.94 & 2.57 & 15.2 & 5.5 & 
1.8 $\pm$ 0.3 & 2002 Apr 25 & 3309\\
FTM1022$+$1929 & 10 22 29.40 & $+$19 29 39.00 & 0.41 & 2.34 & 15.2 & 4.4 & 
0.5 $\pm$ 0.3 & 2004 Mar 28 & 6607\\
FTM1036$+$2828 & 10 36 33.50 & $+$28 28 21.60 & 1.76 & 2.08 & 15.3 & 4.9 & 
0.5 $\pm$ 0.3\tablenotemark{2} & 2002 May 07 & 4136\\
\enddata

\tablenotetext{1}{The Galactic absorption was obtained from Skyview's nH 
survey, based on Dickey \& Lockman 1990, \araa 28, 215.}

\tablenotetext{2}{These objects have continuum reddening estimates. For the 
remaining objects, the Balmer decrement was used to derive $E(B-V)$.}

\tablecomments{Units of right ascension are hours, minutes, and seconds, and 
units of declination are degrees, arcminutes, and arcseconds. We use FIRST 
coordinates, which are in J2000 epoch.}

\end{deluxetable}

\clearpage

\begin{deluxetable}{cccccccc}
\tablewidth{0pc}
\tabletypesize\footnotesize
\tablecaption{X-ray parameters \label{ciao}}
\tablehead{
\colhead{} & \colhead{} & \colhead{} & \multicolumn{3}{|c|}{Xray counts} & 
\colhead{$f_x$ (broad)} & \colhead{$L_x$ (0.5 - 10.0 keV)} \\
\colhead{Source} & \colhead{R.A.} & \colhead{Dec.} & 
\multicolumn{1}{|c} {total} & \colhead{soft} & \multicolumn{1}{c|} {hard} & 
\colhead{(erg/$cm^2$/s)} & \colhead{(erg/s)} 
}
\startdata
FTM0134 & 01:34:35.66 & $-$09:31:02.8 & 95 $\pm$ 12 & 52 $\pm$ 9 & 40 $\pm$ 8 
& 7.57 $\times 10^{-13}$ & 1.10 $\times 10^{46}$ \\
FTM0729 & 07:29:10.33 & $+$33:36:33.9 & 6.0 $\pm$ 3.7 & 0 & 5.0 $\pm$ 3.6 & 
3.04 $\times 10^{-14}$ & 1.01 $\times 10^{44}$ \\
FTM0738 & {\it not} & {\it detected} & - & - & - & 
$\approx 5 \times 10^{-15}$ & - \\
FTM0830 & 08:30:11.15 & $+$37:59:51.8 & 790 $\pm$ 30 & 400 $\pm$ 20 & 390 
$\pm$ 20 & 7.90 $\times 10^{-13}$ & 3.42 $\times 10^{44}$ \\
FTM0841 & 08:41:04.98 & $+$36:04:50.2 & 5.4 $\pm$ 3.8 & 3.0 $\pm$ 3.0 & 3.0 
$\pm$ 3.0 & 5.17 $\times 10^{-15}$ & 4.52 $\times 10^{42}$ \\
FTM0904 & 09:04:50.53 & $-$01:45:24.7 & 6.0 $\pm$ 3.7 & 3.0 $\pm$ 3.1 & 3.0 
$\pm$ 3.1 & 6.89 $\times 10^{-15}$ & 2.61 $\times 10^{43}$ \\
\\
FTM0906 & 09:06:51.53 & $+$49:52:35.8 & 40 $\pm$ 8 & 26 $\pm$ 7 & 13 $\pm$ 5 
& 6.12 $\times 10^{-14}$ & 7.35 $\times 10^{44}$ \\
FTM0915 & 09:15:01.72 & $+$24:18:11.9 & 68 $\pm$ 10 & 23 $\pm$ 7 & 48 $\pm$ 8 
& 1.92 $\times 10^{-13}$ & 4.72 $\times 10^{44}$ \\
FTM1004 & 10:04:24.88 & $+$12:29:22.4 & 41 $\pm$ 9 & 22 $\pm$ 7 & 26 $\pm$ 10 
& 6.74 $\times 10^{-14}$ & 2.31 $\times 10^{45}$ \\
FTM1012 & 10:12:30.48 & $+$28:25:26.0 & 10 $\pm$ 5 & 5.4 $\pm$ 3.8 & 5.0 
$\pm$ 3.6  & 2.69 $\times 10^{-14}$ & 8.73 $\times 10^{44}$ \\
FTM1022 & 10:22:29.39 & $+$19:29:39.1 & 11 $\pm$ 5 & 3.9 $\pm$ 3.7 & 7.4 
$\pm$ 4.2 & 1.45 $\times 10^{-14}$ & 6.28 $\times 10^{42}$ \\
FTM1036 & 10:36:33.55 & $+$28:28:21.2 & 48 $\pm$ 8 & 34 $\pm$ 7 & 19 $\pm$ 6 
& 9.70 $\times 10^{-14}$ & 1.36 $\times 10^{45}$ \\

\enddata

\tablecomments{The flux given for the not detected source is the background 
flux and is considered an upper limit. Coordinates listed are in J2000 epoch 
and are for the Chandra X-ray data. Broad band is 0.2 - 10.0 keV, the soft 
band spans the 0.2 - 2.0 keV range, while the hard band energy range is 2.0 - 
10.0 keV. The luminosity is given in the 0.5 - 10.0 keV rest frame.}

\end{deluxetable}

\clearpage

\begin{deluxetable}{cccc}
\tablewidth{0pt}
\tabletypesize\footnotesize
\tablecaption{Hardness ratios
\label{hr}}
\tablehead{
Source & \colhead{HR} & \colhead{HR} & \colhead{corrected HR} \\
 & \colhead{0.2 - 2.0 and 2.0 - 10.0 keV} & 
\colhead{0.5 - 2.0 and 2.0 - 7.0 keV} & 
\colhead{0.4 - 4.0 and 4.0 - 20.0 keV} \\
 & \colhead{observed frame} & \colhead{observed frame} & 
\colhead{rest frame}
}
\startdata
FTM0134$-$0931 & $-$0.12 $\pm$ 0.13 & $-$0.06 $\pm$ 0.13 & 
$+$0.54 $\pm$ 0.12 \\
FTM0729$+$3336 & $+$1.00 $\pm$ 0.72 & $+$1.00 $\pm$ 0.72 & 
$+$1.00 $\pm$ 0.40 \\
FTM0738$+$2750 &  ---               &  ---               & --- \\
FTM0830$+$3759 & $-$0.02 $\pm$ 0.04 & $-$0.02 $\pm$ 0.04 & 
$-$0.34 $\pm$ 0.04 \\
FTM0841$+$3604 & $+$0.00 $\pm$ 0.71 & $+$0.00 $\pm$ 0.71 & 
$+$0.00 $\pm$ 0.71 \\
FTM0904$-$0145 & $+$0.00 $\pm$ 0.73 & $+$0.00 $\pm$ 0.73 & 
$+$0.00 $\pm$ 0.73 \\
FTM0906$+$4952 & $-$0.34 $\pm$ 0.20 & $-$0.34 $\pm$ 0.20 & 
$-0.16$ $\pm$ 0.21 \\
FTM0915$+$2418 & $+$0.36 $\pm$ 0.15 & $+$0.36 $\pm$ 0.15 & 
$+$0.26 $\pm$ 0.15 \\
FTM1004$+$1229 & $+$0.10 $\pm$ 0.23 & $+$0.10 $\pm$ 0.23 & 
$+$0.55 $\pm$ 0.26 \\
FTM1012$+$2825 & $-$0.04 $\pm$ 0.50 & $-$0.04 $\pm$ 0.50 & 
$-$0.04 $\pm$ 0.50 \\
FTM1022$+$1929 & $+$0.32 $\pm$ 0.50 & $+$0.34 $\pm$ 0.45 & 
$-$0.21 $\pm$ 0.43 \\
FTM1036$+$2828 & $-$0.27 $\pm$ 0.17 & $-$0.30 $\pm$ 0.17 & 
$-$0.02 $\pm$ 0.18 \\
\enddata

\tablecomments{The first HR uses 0.2 - 2.0 keV as the soft and 2.0 - 10.0 keV 
as the hard band to obtain the most photons in each band. The second HR is 
the HR typically used in Chandra surveys, 0.5 - 2.0 keV for the soft and 
2.0 - 7.0 keV for the hard band. The last HR is corrected for redshift, so we 
used the bands of the first HR at z = 1 or 0.4 - 4.0 keV for the soft and 
4.0 - 20.0 keV for the hard band in the quasar rest frame.}

\end{deluxetable}

\clearpage

\begin{deluxetable}{cccc}
\tablewidth{0pt}
\tablecaption{{\it XSPEC} fitting information to an absorbed power law 
\label{xspec}}
\tablehead{
 & \colhead{$N_H$} & & \colhead{corrected $L_x$ (erg/s)} \\
\colhead{Source} & \colhead{($\times 10^{22} cm^{-2}$)} & \colhead{$\Gamma$} 
& \colhead{(0.5-10 keV rest frame) }
}
\startdata
FTM0134$-$0931$^a$ & 5.9 $\pm$  2.8 & 1.8 $\pm$ 0.5 & 2.4 $\times 10^{46}$ \\
FTM0134$-$0931$^b$ & 1.4 $\pm$  0.6 & 1.8 $\pm$ 0.5 & 2.4 $\times 10^{46}$ \\
FTM0830$+$3759 & 2.7 $\pm$ 0.2 & 2.9 $\pm$ 0.1 & 2.1 $\times 10^{45}$ \\
FTM0906$+$4952 & 0.6 $\pm$ 0.9 & 1.5 $\pm$ 1.0 & 6.3 $\times 10^{44}$ \\
FTM0915$+$2418 & 6.7 $\pm$ 4.3 & 2.5 $\pm$ 1.2 & 2.3 $\times 10^{45}$\\
FTM1004$+$1229 & 28 $\pm$ 25 & 2.3 $\pm$ 1.3 & 8.6 $\times 10^{45}$\\
FTM1036$+$2828 & 3.8 $\pm$ 3.4 & 2.3 $\pm$ 1.3 & 3.5 $\times 10^{45}$ \\
\enddata

\tablecomments{$^a$ Absorption at quasar redshift, 
$^b$ Absorption at lens redshift. The luminosity is corrected for absorption 
and is given in rest frame band 0.5 - 10 keV. Note that for FTM0906$+$4952 the 
luminosity actually decreases because it has a flatter spectral slope than 
the assumed $\Gamma$ = 2 in Table \ref{ciao} and a relatively small 
absorption.}

\end{deluxetable}

\end{document}